\documentclass[sigconf]{acmart}

\begin{document}
\fancyhead{}
\title{Towards a Data Centric Approach for the Design and Verification of Cryptographic Protocols\footnotemark[2]}

\author{Luca Arnaboldi}
\authornote{Both authors contributed equally to this work}
\affiliation{%
  \institution{Newcastle University}
}
\email{l.arnaboldi@ncl.ac.uk}
\author{Roberto Metere}
\authornotemark[1]
\affiliation{%
  \institution{Newcastle University and\and The Alan Turing Institute}
}
\email{roberto.metere@ncl.ac.uk}

\begin{abstract}
 We propose MetaCP, a Meta Cryptography Protocol verification tool, as an automated tool simplifying the design of security protocols through a graphical interface.
 The graphical interface can be seen as a modern editor of a non-relational database whose data are protocols.
 The information of protocols are stored in XML, enjoying a fixed format and syntax aiming to contain all required information to specify any kind of protocol.
 This XML can be seen as an almost semanticless language, where different plugins confer strict semantics modelling the protocol into a variety of back-end verification languages.
 In this paper, we showcase the effectiveness of this novel approach by demonstrating how easy MetaCP makes it to design and verify a protocol going from the graphical design to formally verified protocol using a Tamarin prover plugin.
 Whilst similar approaches have been proposed in the past, most famously the AVISPA Tool, no previous approach provides such as small learning curve and ease of use even for non security professionals, combined with the flexibility to integrate with the state of the art verification tools.
\end{abstract}


\renewcommand\footnotetextcopyrightpermission[1]{}
\setcopyright{none}
\settopmatter{printacmref=false}

\maketitle

\renewcommand{\thefootnote}{\fnsymbol{footnote}}
\footnotetext[2]{\em Final version to appear in the Proceedings of the 26th ACM SIGSAC Conference on Computer and Communications Security (CCS'19) - Posters \& Demo section.}

\section{Introduction}

\begin{figure}[h]
  \caption{A typical approach to the verification of a protocol (top) versus our centralised approach (bottom).}
  \label{fig:designprocess}
	  \centering\includegraphics[width=\columnwidth]{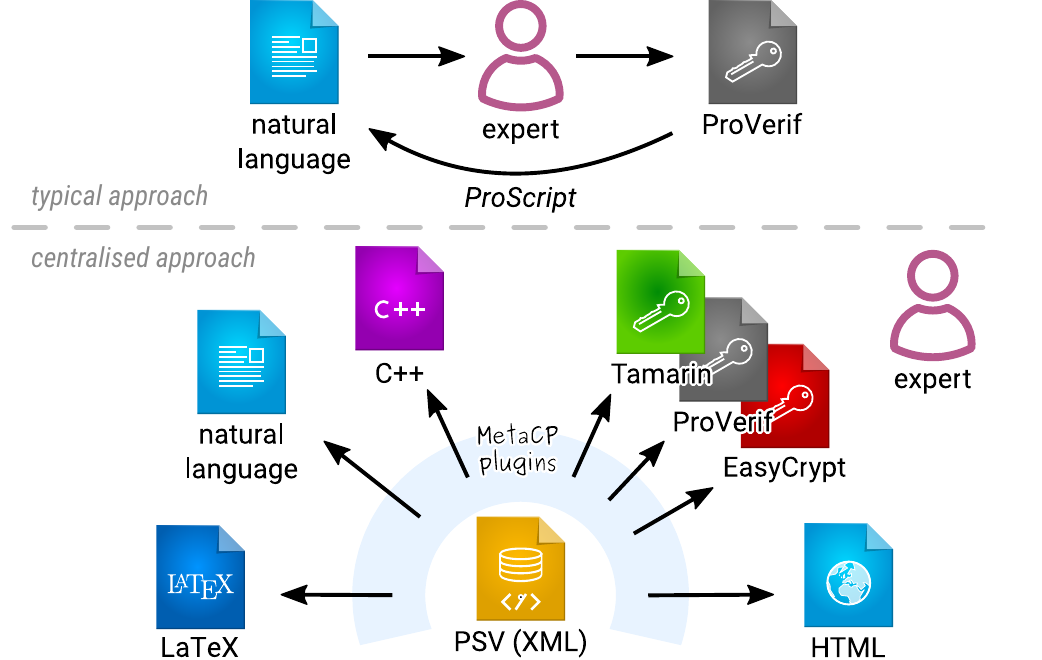}
\end{figure}

When communication travels through insecure channels, as it does many times per day in the Internet, high level of assurance of security of sensible information becomes of paramount importance.
The rules of communications between parties are dictated by protocol implementations.
In this context, attackers and protocols continuously evolve in an adversarial game where the protocols improve to be secure against the attackers.
Before implementing a protocol, we usually desire to ensure that the design is not flawed.
For this purpose, the application of formal methods to verify protocols has proved itself to be very effective.
Protocol verification languages such as ProVerif~\cite{blanchet2001efficient}, Tamarin~\cite{meier2013tamarin}, EasyCrypt~\cite{barthe2011computer} 
and many more, have been proposed to provide aided semi automatic verification of protocols with excellent success ~\cite{metere2017automated,metere2018speke,arnaboldi2019AceOAuth,basin2018formal}; however, there is a huge disconnection between these languages and the actual protocols. 
For those who are not experts in a specific language, interpreting a specification of a protocol written in English and modelling it into a formal tool is a difficult task. 
Even harder is the act of going back from the model to the original specification as the model may contain less information than the specification.
What is desirable instead is a single structured interface that is easy to understand, visualise, and that can automatically transform a design into a variety of back-end verification options, ensuring correctness of design and providing consistency.
We illustrate this data-centric approach in~Fig.~\ref{fig:designprocess}.
In particular, in our approach the specifications are natural language agnostic and stored in a database containing all required information; the semantics of such information are determined only when a spoken, visual, or formal language specification is created from it.
Having this central source of information would be beneficial to link all the different (and disjointed) semantics we usually adopt in different contexts.
In addition, the protocols could also enjoy a much higher degree of rigour they allegedly lack with the current workflow.

To showcase the effectiveness of our approach, we implement and present the promising results we obtained by using our tool MetaCP.
MetaCP is a tool that enables the designers of security protocols to be guided through their work, and allowing them to rely on automatic reasoning of supported security properties at each step of the design and verification process.

\section{Design of secure protocols}

The design of a protocol almost always follows the same three intertwining steps; 
i) a graphical representation of the protocol is sketched out, representing the involved parties and the flow of the protocol; 
ii) message contents and mathematical requirements are set out; 
iii) security goals are decided upon, often not formally but rather written in natural language.
The processes of designing the protocol and employing a formal verification tool are often detached and done at separate stages by different people. 
The state of the art tools to perform formal security evaluation, such as Tamarin and ProVerif, improved in the past decades and enjoy a wide spread use. 
However, these tools do not provide any means to relate to the whole design process, impacting the usability and effectiveness of the evaluations.
As it stands, it is very difficult for the casual user or even for the security professional, to ascertain the truthfulness of the analysis, and how it relates to the original protocol.

One further negative aspect of the current process is that each tool uses a separate syntax requiring a designer to be able to interpret various different idioms for verification. 
This is an unfortunate circumstance of the various tools giving different results for the same protocol~\cite{metere2018speke}.
Witnessing the sensibility of researchers about this problem, namely the project ``Automated Validation of Internet Security Protocols
and Applications'' (AVISPA)~\cite{armando2005avispa} has attempted to unify the verification, by presenting a single language of specification and automating the translation into various back-end tooling. 
The proposed approach based itself on the assumption that most people would be familiar with AnB notation~\cite{caleiro2006semantics}.
To this end, they extended and formalised several aspects of this notation~\cite{caleiro2006semantics} which eventually resulted in HLPSL.
From this language the tool is able to automatically convert into several verification back-ends to perform the analysis.
Even with the integration of multiple verification options, research shows that protocols found to be secure by AVISPA were later found flawed in more modern tools~\cite{metere2018speke}.
This demonstrates the need for ease of extension to integrate more tooling in the back-end verification.

Another attempt that the literature proposes to standardise the way in which protocols are designed is ProScript ~\cite{kobeissi2017automated}, here, a new high level language for the specification of security protocols is proposed.
ProScript is able to automatically interpret from the high level specification to applied PI calculus, verifiable in ProVerif and CryptoVerif. 
Their approach is very much inline with our desired outcome, however their new definition language has the major downside that it doesn't allow for much to be expressed, somewhat limiting what you can describe as well as being restricted to the usage with the applied PI calculus based tools.

Both AVISPA and ProScript certainly make it easier to verify and design a protocol; in contrast, their reliance on a single semantic of translation makes it difficult to expand to newer tools with different semantics.
We strongly believe that our approach of using only basic semantics and enforcing a specific one in the target language generation process is much more extendable and promising.
This is because this approach does not pretend to substitute the typical human process of modelling the reality to a mathematical representation, but rather to aid it.

\begin{figure}[h]
  \caption{Design Interface for MetaCP}
  \label{fig:metacpinterface}
  \centering\includegraphics[width=\columnwidth]{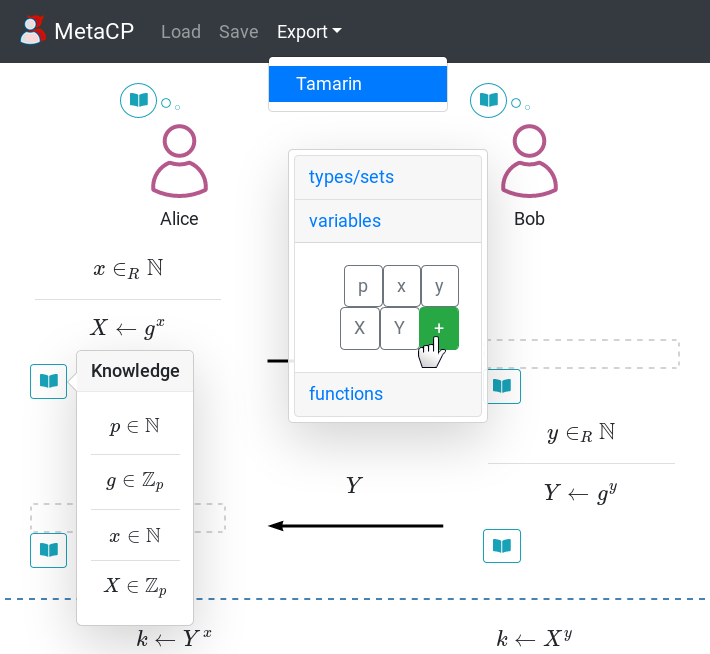}
\end{figure}

What MetaCP proposes is a graphical design interface, as illustrated in~Fig.~\ref{fig:metacpinterface}, that allows a protocol to be drawn out easily through a drag-n-drop interface.
The interface can be seen as a database graphical editor able to let the user specify variables, functions, message flow, and equational theories; then it lets the user embed all the possible knowledge required to describe a protocol, going far beyond what is possible in symbolic model checkers, and providing more rigour to common natural language specification.

\section{From specification to Verification}

Once the protocol is specified using the MetaCP tool the graphical representation is saved as a structured XML containing all information required to describe the protocol; this structured XML might be seen as a basic language with syntax and very little semantics.
The structured XML can be viewed as a database of Protocol Specification and Verification (PSV).
To allow non-experts to work with it, we allow the creation and modification of protocols through a flexible and easy to use graphical interface; editiing through the PSV is also allowed and can be visualised back into the interface.
We remark that this structure has little to no semantics merely aimed to provide information for the underlying plugins.
The reasoning behind this is that various tools work very differently, whilst one semantic might work very well to translate into one tool it would fail to capture the requirements of another.

With the continuous evolution of new tools and models,  
a tool that scales to incorporate these approaches is very important.
To allow for this scalability, a plugin system is devised to work with the PSV, where each plugin enforces the desired semantics of the target language from the underlying PSV.
Since modelling a protocol is done in the first place as an effective, arbitrary interpretation, one of the goals is usually to show that the model captures the properties required.
This task is delegated to experts in the target language.

By having a PSV, you can store much more information than what is required by symbolic model checkers thus,
we are not limited to the automated conversion of the protocol to a symbolic modelling language.
Translating plugins may even be added for the automated conversion to natural language to create English written specifications, and even to program code such a C++.
The ability to store further relations and information about the protocol allows for syntax checking and errors in the design; in particular, we emphasise that this {\em simple} aid is currently unavailable in successful verification tools such as Tamarin.
This support is however currently basic in MetaCP and we reserve its expansion and improvement for future development.

\section{Experiment results}

The current development state of MetaCP supports a single plugin specifically intended to replicate the semantics of Tamarin.
Tamarin uses multiset rewriting rules to specify the protocol and provides an efficient semi-automatic verification engine. 
The plugin provides a fully automated protocol-agnostic interpreter from PSV code to Tamarin code. 
To demonstrate the efficacy of our plugin, we designed three protocols: 
i) Diffie-Hellman key exchange (DHKE), a popular protocol to the majority of security experts;
ii) Needham-Schroeder (NSP, asymmetric), a fundamental example of design errors that can be easily patched to meet security, and finally
iii) the Needham-Schroeder-Lowe (NSLP), the fixed version of NSP.
Our intent was to compare the results from the automatically generated models to the results of the models manually generated available as official Tamaring examples.
We would like to remark that our plugin is merely \textit{a} plugin for Tamarin; additional plugins targeting the same language are allowed, and whilst that is out of the scope of this current work it showcases the flexibility of this approach.
\begin{table}[t]
\caption{Results of experiment using Tamarin. We compare the code automatically produced by MetaCP with the official examples of the Tamarin tool.}
\label{tab:tableResults}
\begin{tabular}{rcccc}
\toprule
  & \multicolumn{2}{c}{lines of code} & \multicolumn{2}{c}{verification time} \\
  & manual & auto & manual & auto \\
  \midrule
  Diffie-Hellman       &  -  &  85 &  -     & 0.23 s \\
  Needham-Schroeder      & 146 & 118 & 2.37 s & 1.78 s \\
  Needham-Schroeder-Lowe &  83 &  90 & 1.90 s & 1.42 s \\
  \bottomrule
\end{tabular}
\end{table}
Using the interface and some manual intervention, we quickly and automatically generated the three protocols, then we automatically exported them to Tamarin.
The exported models were well formed and passed the correctness lemmas.
In addition to the correctness, we manually added other security goals to compare to the provided files (a simple copy/paste was sufficient). 

In the comparison between the automatically generated and the sample Tamarin models, we could successfully show the exact same attacks as well as the same proofs of security.
Even without the currently unimplemented security goals, our tool significantly reduces the required work as we integrated the security goals into the generated output at a fraction of the time the full writing out would require - not mentioning the knowledge and expertise needed to manually write Tamarin code.
In our experimentation, we found that some protocols in Tamarin need extra help to reduce the state space: the intuitive cryptographic decomposition as expressed in PSV is not optimally interpreted and slows down the verification significantly, whilst breaking down a message led to its quick acceleration.
Finally, adding two lines as well as the lemmas, we were able to exactly match the security results of the official Tamarin examples written by experts, as illustrated in Table~\ref{tab:tableResults}.

\section{Future Work}

We provide a high level design interface to design and specify security protocols, embellished by an automated translation from design to knowledge base to formal specification in Tamarin.
Although this work is still immature, its initial results are definitely very promising: we were able to design and verify protocols much more efficiently and less prone to errors and imprecision that may have occurred had we done it manually.
In particular, a simple sketch of the DHKE protocol could be written in less than ten minutes, including its correct formalisation and verification in Tamarin.
Future work would involve enriching the tools with novel plugins developed by us targeting the most common protocol tools, including diverse languages such as EasyCrypt.
We also aim to make the tool open source for anyone to use and extend to include their own plugin and even different plugin options for the same backends.

As a farsighted final conclusion, it is a worthwhile exercise to attempt to standardise the methodology in which future (and current) protocols are designed.
Formally verified models confer to protocols the reliability required to focus on the problems that raise at the implementation level.
We see the data centric methodology of MetaCP as a promising example towards this approach.

\section*{Acknowledgement}
This research is supported by The Alan Turing Institute and an Innovate UK grant to Newcastle University through the e4future project, as well as Arm Ltd. and EPSRC under grant EP/N509528/1.

\bibliographystyle{ACM-Reference-Format}
\bibliography{bibliography}

\end{document}